\documentclass[12pt]{article}
\usepackage{graphicx,setspace}
\usepackage{caption}
\newcommand\pubnumber{NuPhys2018-Brunetti}
\newcommand\pubdate{March 22, 2019}
\def\padova{Istituto Nazionale di Fisica Nucleare - Sezione di Padova\\
Via Marzolo 8, 35131 Padova, Italy}
\def\collab{\footnote{F. Acerbi
, G. Ballerini
, M. Bonesini
, A. Branca
, C. Brizzolari
, M. Calviani
, S. Carturan
, M.G. Catanesi
, S. Cecchini
, F. Cindolo
, G. Collazuol
, E. Conti
, F. Dal Corso
, G. De Rosa
, C. Delogu
, A. Falcone
, B. Goddard
, A. Gola
, R.A. Intonti
, C. Jollet
, V. Kain
, B. Klicek
, Y. Kudenko
, M. Laveder
, A. Longhin
, L. Ludovici
, L. Magaletti
, G. Mandrioli
, A. Margotti
, V. Mascagna
, N. Mauri
, A. Meregaglia
, M. Mezzetto
, M. Nessi
, A. Paoloni
, M. Pari
, E. Parozzi
, L. Pasqualini
, G. Paternoster
, L. Patrizii
, C. Piemonte
, M. Pozzato
, M. Prest
, F. Pupilli
, E. Radicioni
, C. Riccio
, A.C. Ruggeri
, G. Sirri
, M. Soldani
, M. Stipcevic
, M. Tenti
, F. Terranova
, M. Torti
, E. Vallazza
, F. Velotti
, M. Vesco
, L. Votano
}}

\def\support{\footnote{This project has received funding from the European Research Council (ERC) under the European Unions Horizon 2020 research and 
innovation programme (grant agreement N. 681647).}}
\def\Title#1{\begin{center} {\Large #1 } \end{center}}
\def\Author#1{\begin{center}{ \sc #1} \end{center}}
\def\Address#1{\begin{center}{ \it #1} \end{center}}

\newcommand\pubblock{\rightline{\begin{tabular}{l} \pubnumber\\
         \pubdate  \end{tabular}}}
\newenvironment{Abstract}{\begin{quotation}  }{\end{quotation}}
\newenvironment{Presented}{\begin{quotation} \begin{center} 
             PRESENTED AT\end{center}\bigskip 
      \begin{center}\begin{large}}{\end{large}\end{center} \end{quotation}}

\def\beq{\begin{equation}}
\def\eeq#1{\label{#1}\end{equation}}
\def\eeqn{\end{equation}}

\def\beqa{\begin{eqnarray}}
\def\eeqa#1{\label{#1}\end{eqnarray}}
\def\eeqan{\end{eqnarray}}

\let\bar=\overbar

\def\etal{{\it et al.}}

\def\O{{\cal O}}

\def\Dslash{\not{\hbox{\kern-4pt $D$}}}
\def\dslash{\not{\hbox{\kern-2pt $\del$}}}

\def\msb{{\bar{\ssstyle M \kern -1pt S}}}

\begin{document}
\begin{titlepage}
\pubblock

\vfill
\Title{The ENUBET Beamline}
\vfill
\Author{Giulia Brunetti\support}
\Address{\padova}
\begin{center} on behalf of the ENUBET collaboration\collab \end{center}
\vfill
\begin{Abstract}
The ENUBET ERC project (2016-2021) is studying a narrow band neutrino beam where lepton production can be monitored at single particle level 
in an instrumented decay tunnel. This would allow to measure $\nu_{\mu}$ and $\nu_{e}$ cross sections with a precision improved by about one order of
magnitude compared to present results.
 
In this proceeding we describe a first realistic design of the hadron beamline based on a dipole coupled to a pair of quadrupole triplets along 
with the optimisation guidelines and the results of a simulation based on G4beamline. 
A static focusing design, though less efficient than a horn-based solution, results several times more efficient than originally expected. 
It works with slow proton extractions reducing drastically pile-up effects in the decay tunnel and it paves the way towards a time-tagged neutrino beam.
On the other hand a horn-based transferline would ensure higher yields at the tunnel entrance. The first studies conducted at CERN to implement 
the synchronization between a few ms proton extraction and a horn pulse of 2-10~ms are also described.

\end{Abstract}
\vfill
\begin{Presented}
NuPhys2018, Prospects in Neutrino Physics\\
Cavendish Conference Centre, London, UK, \\
December 19--21, 2018
\end{Presented}
\vfill
\end{titlepage}
\def\thefootnote{\fnsymbol{footnote}}
\setcounter{footnote}{0}
%
\section{ENUBET (Enhanced NeUtrino BEams from \\kaon Tagging)}

Neutrino experiments are now limited by the knowledge of the initial fluxes, the current achievable precision in the absolute cross section measurements
is \O(5-10\%). A dedicated facility based on conventional accelerator techniques and existing infrastructures designed to address this problem would 
impact the entire field of neutrino oscillation physics. The ENUBET facility \cite{epjc,eoi,loi} addresses simultaneously the two most important 
challenges of the next generation neutrino experiments: a superior control of the flux and flavour composition at source and a high level of tunability 
and precision in the selection of the energy of the outgoing neutrinos.
At present the flux of $\nu_{\mu}$ beams is not directly measured but relies on detailed simulations of the neutrino beamline and on extrapolations of 
target hadro-production data. Moreover, most of the next generation oscillation experiments will measure $\nu_{e}$ appearance at the far detector.
By improving the precision on $\nu_{\mu}$ and $\nu_{e}$ cross sections by about one order of magnitude ENUBET results would be of great value for 
current (NOvA, T2K) and next generation long-baseline experiments (DUNE, Hyper-Kamiokande).

The ENUBET proposal \cite{epjc} was motivated by the idea of developing ``monitored neutrino beams": a facility where the only source of $\nu_{e}$ is the three-body 
semileptonic decay of kaons: $K^{+} \rightarrow \pi^{0}e^{+}\nu_{e}$ ($K_{e3}$). The goal is to build a detector capable of identifying the positrons 
from $K_{e3}$ decays while operating in the harsh environment of a conventional neutrino beam decay tunnel. 

The ENUBET beamline will allow performing $\nu_{\mu}$ cross section studies with a narrow band beam where the neutrino energy is known a priori 
with 10\% uncertainty and $\nu_{e}$ cross section measurement with 1\% precision with a monitored neutrino beam where the positrons from $K_{e3}$ decays 
are monitored at single particle level by the calorimeters instrumenting the decay tunnel \cite{calo}. 

This can be achieved using conventional magnets by maximising the number of $K^{+}$ and $\pi^{+}$ at tunnel entrance, by minimising the total length of 
the transferline to reduce kaon decay losses and by keeping under control the level of background transported. Momentum and charge-selected hadrons 
($K^{+}$, $\pi^{+}$) being injected in the instrumented decay tunnel need to be collimated enough such that any undecayed meson is capable of escaping 
the region without hitting the tagger inner surface: this allows not to swamp the instrumentation with excessive particle rates and to limit the 
monitoring to the decay products of $K$ decays. Furthermore it is very important to tune the shielding and the collimators to minimise any beam induced 
background in the decay region.
The beamline presented here is composed by a short ($\sim$20~m) transferline followed by a 40~m long decay tunnel.
The hadron beam considered has a reference momentum of 8.5 GeV/c with a momentum bite of 10\%.

The proton interactions in the target are simulated with FLUKA, we have considered various proton drivers (400 GeV, 120 GeV and 30 GeV protons) and 
target designs. The results reported in this document refer to 400 GeV protons and a beryllium target 110~cm long with a 3~mm diameter. 
The optic optimization is performed with TRANSPORT to match the ENUBET specifications for momentum bite and beam envelope. The beam components and 
lattice are then implemented in G4Beamline that fully simulates particle transport and interactions. 

We have considered two possibile beamlines: the first one makes use of a focusing horn placed between the ENUBET target and the following transferline 
(``horn-based transferline") while in the second one the transferline quadrupoles are placed directly downstream the target (``static transferline").

Here we describe more in detail the static design, whose performance turned out to be significantly better than early estimates reported in 
the ENUBET proposal \cite{epjc} and it offers several advantages in terms of cost, simplification of technical implementation and performance of 
particle identification.  Moreover a static transferline would pave the way to the so-called tagged-beams. A ``tagged neutrino beam" is a facility 
where the neutrino is uniquely associated with the other particles of the parent kaon.
Since in the static focusing system the proton extraction can last up to several seconds, the instantaneous rates of particles hitting 
the decay tunnel walls is reduced by about two orders of magnitude compared with the horn option. In the ENUBET static option the time between two 
$K_{e3}$ decays is 1.3~ns. A neutrino interaction in the detector can thus be time linked with the observation of its associated lepton in the decay 
tunnel: this has never been perfomed in any neutrino experiment and would represent a major breakthrough in experimental neutrino physics. 

Results obtained for the horn-based transfer line are presented as well and related studies are also being pursued due to the remarkable 
$\nu$ fluxes that can be achieved.

\section{Static transferline}
\begin{figure}[!h]
\centering
\includegraphics[width=\textwidth]{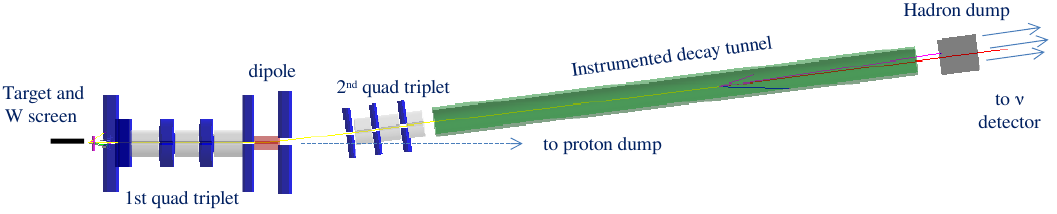}
\caption{
Schematics of the ENUBET beam in the static focusing option.}
\label{fig:tl}
\end{figure}

The static configuration is very promising since it allows to perform the focusing using DC operated devices (unlike pulsed magnetic horns) compatible 
with a traditional slow extraction of several seconds. The ENUBET beam (see Figure \ref{fig:tl}) is a conventional narrow band beam where, unlike most of the current beams, the decay 
tunnel is not located in front of the focusing system and the proton extraction length is slow (2~s). The best configuration achieved consists in a 
quadrupole triplet followed by a dipole that provides a 7.4$^{\circ}$ bending angle and by another quadrupole triplet.

\begin{figure}[!h]
\centering
\includegraphics[width=\textwidth]{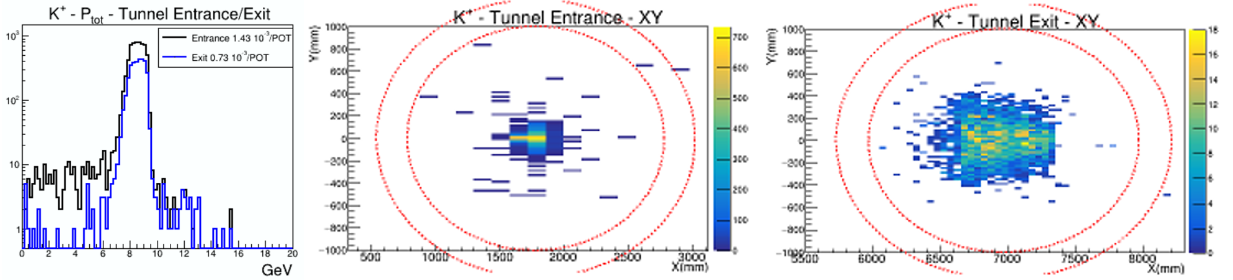}
\caption{
Left: momentum distribution of $K^{+}$ entering/exiting the decay tunnel. Right: XY profile of the $K^{+}$ beam at tunnel entrance and exit.}
\label{fig:plots}
\end{figure}

Particles produced by proton interactions in the target are focused, momentum selected and transported to the tunnel entrance. Non-interacting protons are stopped in a proton beam dump. 
Off-momentum particles reaching the decay tunnel are mostly low energy particles coming from interactions in the collimators and other beamline 
components together with muons that cross absorbers and collimators. At 8.5 Gev/c we expect $\sim$50\% of $K^{+}$ to decay in a 40~m long tunnel.
The rate of background particles is several order of magnitude smaller than present beams and the instrumentation located in the decay tunnel can 
monitor lepton production at single particle level. Figure \ref{fig:plots} shows the momentum distrbution as well as the XY profile of 
$K^{+}$ entering/exiting the decay tunnel.

The length of the decay tunnel is optimized in order to have $K_{e3}$ decays as the only $\nu_{e}$ source: electron neutrinos from decay in flight of 
kaons represent $\sim$97\% of the overall $\nu_{e}$ flux. The positrons from three body decays are emitted at large angles and hit the instrumented 
walls of the tunnel before exiting. 
The identification of particle hitting the tunnel walls is performed by longitudinally segmented calorimeters \cite{calo}, positrons are separated from 
photons using a photon veto made of plastic scintillators tiles located just below the innermost layer.

The static beamline transports at the tunnel entrance 19~10$^{-3}$ $\pi^{+}$/POT and 1.4~10$^{-3}$ $K^{+}$/POT in [6.5$\div$10.5 GeV/c] range, improving 
by 4 times the kaon yield with respect to the first estimate reported in \cite{epjc} and requiring about 4.5~10$^{19}$ POT at CERN SPS to carry 
out both $\nu_{\mu}$ and $\nu_{e}$ cross section programs. 
An additional advantage of the static solution is the possibility to directly monitor the rate of muons from $\pi^{+}$ decays after the hadron 
dump, it cannot be done for the higher rates in the horn-based solution but since it is reduced by two order of magnitude in the static option the 
$\nu_{\mu}$ flux can be monitored with the same level of precision of $\nu_{e}$.

\section{Horn-based transferline}

\begin{figure}[!h]
\centering
\includegraphics[width=\textwidth]{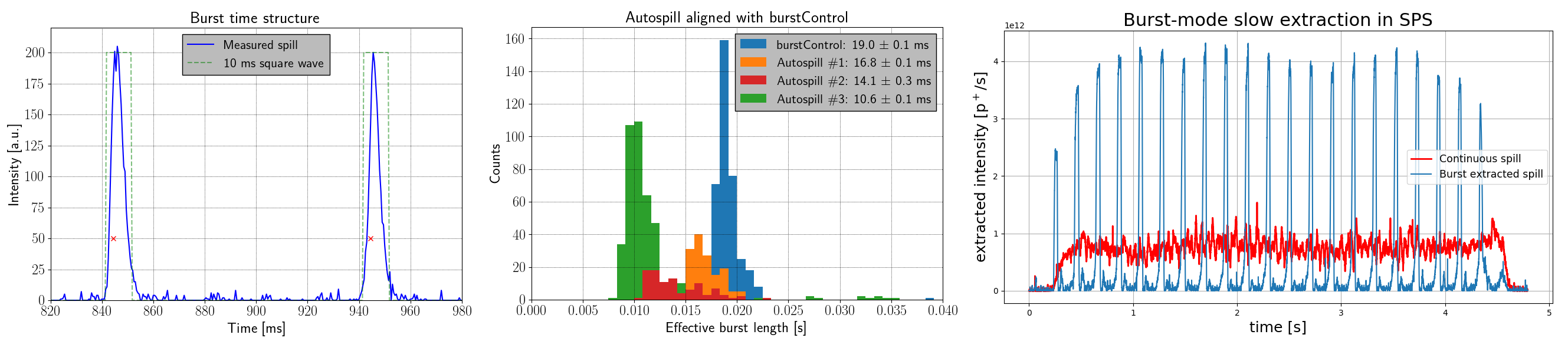}
\caption{
Left: Algorithm of feed-forward implemented and capable of optimizing the burst length towards 10~ms. Center: Proof that the algorithm is capable to 
reduce the burst length form $\sim$20~ms to $\sim$10~ms (10.6~ms) in 3 interactions. Right: burst-extraction over a whole SPS spill (CERN-BE-OP-SPS, 
F.Velotti, M.Pari, V.Kain, B.Goddard).}
\label{fig:cern}
\end{figure}

In the horn-based solution a magnetic horn is placed between the target and the following transferline. This horn needs to be pulsed for 2-10~ms and 
cycled at several Hz during the accelerator flat-top. The studies concerning the proton extraction scheme (``burst-mode extraction") to combine a few ms 
proton extraction with 2-10~ms horn pulses are on-going at CERN. As presented in Figure \ref{fig:cern} we could already confirm the 
proof-of-concept of feed-forward burst spill optimization: the ``Autospill-Burst" algorithm developed leads to a burst length optimization from 20 to 10.6 ms. From this 
benchmark the studies will continue to explore the full simulation and to address remaining issues towards the full operability. 

The flux produced at the tunnel entrance is 4-5 times larger than in the static option: at the SPS we expect 77~10$^{-3}$ $\pi^{+}$/POT and 
7.9~10$^{-3}$ $K^{+}$/POT in [6.5$\div$10.5 GeV/c] range. This represents an improvement factor of 2 in kaon transport with respect to the first estimate 
reported in \cite{epjc}.

\end{document}